\documentclass[prc,twocolumn,showpacs,superscriptaddress,preprintnumbers,amsmath,amssymb,nofootinbib]{revtex4-1}
\usepackage{bm}
\usepackage{dcolumn}
\usepackage{multirow}
\usepackage{ifpdf}
\usepackage{url}

\ifpdf
  \usepackage{graphicx}     
  \usepackage{hyperref}
\else     
  \usepackage[dvipdfmx]{graphicx}     
  \usepackage[dvipdfmx]{hyperref}
\fi
\usepackage{color} 

\newcommand\Tmunu{T^{\mu\nu}}

\newcommand\roundmu{\partial_\mu}

\newcommand\pythia{\textsc{Pythia}}

\def\sNN{\sqrt{s_{NN}}}
 
\hypersetup{
colorlinks=true,
linkcolor=blue,
citecolor=blue,
urlcolor=blue
}

\begin{document}

\title{
Unified description of hadron yield ratios \\
from  dynamical core--corona initialization
}
\author{Yuuka Kanakubo}
\email{y-kanakubo-75t@eagle.sophia.ac.jp}
\affiliation{%
Department of Physics, Sophia University, Tokyo 102-8554, Japan
}

\author{Yasuki Tachibana}
\email{yasuki.tachibana@wayne.edu}
\affiliation{
Department of Physics and Astronomy, Wayne State University, 
Detroit, MI 48201, USA
}
\affiliation{%
Department of Physics, Sophia University, Tokyo 102-8554, Japan
}
\author{Tetsufumi Hirano}
\email{hirano@sophia.ac.jp}
\affiliation{%
Department of Physics, Sophia University, Tokyo 102-8554, Japan
}

\date{\today}

\begin{abstract}
We develop the dynamical core--corona initialization framework 
as a phenomenological description of the formation of quark gluon plasma (QGP)
fluids in high-energy nuclear collisions. Using this framework, we investigate
the fraction of the fluidized energy to the total energy
and strange hadron yield ratios as functions of multiplicity,
and we scrutinize the multiplicity scaling of hadron yield
ratios recently reported by the ALICE Collaboration.
Our results strongly indicate that the QGP  fluids
are partly formed even at the averaged multiplicity for nonsingle diffractive p+p events.
\end{abstract}


\maketitle

\section{Introduction}
Properties of the quark gluon plasma (QGP) have been 
investigated through high-energy nuclear collision experiments 
at the Relativistic Heavy Ion Collider (RHIC) at Brookhaven National 
Laboratory and at the Large Hadron Collider (LHC) at CERN.
The experimental results have brought us a great deal of information about 
properties of the QGP.
The collective behavior of the QGP is well described by 
relativistic hydrodynamics \cite{Heinz:2001xi,Lee:2005gw,Gyulassy:2004zy,Shuryak:2004cy,Hirano:2005wx},
in which the system is postulated to keep local thermal equilibrium during expansion.
Since dissipative effects on expansion are quite small regardless of remarkably 
large expansion rate, the QGP exhibits near-perfect fluidity, which has
led to the concept of strongly coupled/interacting QGP (sQGP) \cite{Lee:2005gw,Gyulassy:2004zy,Shuryak:2004cy}.

Comparisons of the experimental results from heavy-ion collisions 
with those from small colliding systems, such as proton-proton and proton-nucleus collisions,
have been made to see 
how small the QGP droplets can be.
Small colliding systems had been believed to provide us with 
reference data for a long time since the size of the matter
is too small to reach local thermal equilibrium. 
However,
recent experimental data imply QGP 
formation even in small colliding systems (for reviews, see, \textit{e.g.}, Refs.~\cite{Dusling:2015gta,Nagle:2018nvi}). 
One of those data shows
sizable azimuthal anisotropy in final hadron distributions \cite{CMS:2012qk,Abelev:2012ola,Aad:2013fja,Khachatryan:2015waa, Adare:2013piz,Adamczyk:2015xjc,PHENIX:2018lia}, 
which can be interpreted as a result
of the hydrodynamic response of the QGP medium to the initial collision 
geometry \cite{Bozek:2011if,Bzdak:2013zma,Qin:2013bha,Werner:2013ipa,Koop:2015trj,Habich:2015rtj,Shen:2016zpp,Zhao:2017rgg}.
Although this has been discussed actively using relativistic hydrodynamics,
another interpretation is brought about by
the color glass condensate picture,
which gives long-range correlation in rapidity in small colliding systems
 without invoking the QGP formation 
\cite{Dumitru:2010iy,Dusling:2012iga,Dusling:2013qoz,Dumitru:2014yza,Schenke:2015aqa,Lappi:2015cgx,Gotsman:2016whc,Schenke:2016lrs,McLerran:2016snu,Iancu:2017fzn,Dusling:2017dqg,Kovchegov:2018jun,Mace:2018yvl,Zhang:2019dth}.
Thus the origin of collectivity observed in small colliding system still remains to be understood.

Besides the azimuthal anisotropy,
the strangeness enhancement in small colliding systems was
reported by the ALICE Collaboration \cite{ALICE:2017jyt}.
Yield ratios of (multi)strange hadrons to charged pions 
were measured, 
and they exhibited monotonic and continuous increase
 as functions of multiplicity.
Even in small colliding systems, the strange hadron yield ratios
in high-multiplicity events
almost reached the values in heavy-ion collisions.

Strangeness enhancement was proposed a long time ago
to be a signal of the QGP formation in high-energy nuclear collisions
\cite{Rafelski:1982pu,Koch:1986ud}.
Since initial colliding nuclei do not contain the strange quark as the valence one,
yields of strange hadrons would be sensitive to details of reaction dynamics.
If the production rate of strange quarks becomes sufficiently high in thermalized systems,
chemical equilibrium of strange quarks besides up and down quarks is expected to be reached 
within the time scale of nuclear collisions.
In fact, most of the yield ratios of hadrons including multi-strange baryons
have been well reproduced by statistical models \cite{Andronic:2017pug},
which is recognized to be one of the strong 
signals of the QGP formation in heavy-ion collisions
at RHIC and LHC energies.
Therefore the  strangeness enhancement
in high-multiplicity proton-proton and proton-nucleus collisions
reported by the ALICE Collaboration \cite{ALICE:2017jyt} 
strongly indicates the QGP formation
even in such a small colliding system.

This continuous increase of the
strange hadron yield ratio can be interpreted
as follows. 
Final hadrons originate
from two different sources having different hadron production processes.
The values of yield ratios in low-multiplicity
events are
described by
results from string fragmentation, in which the string tension ($\kappa \approx 1$ GeV/fm) 
controls the yield ratios of 
hadrons.
On the other hand, the values of yield ratios in
high-multiplicity
events are
explained by the statistical hadronization approach
where the matter is assumed to be in chemical equilibrium.
In this case, chemical freezeout temperature ($T_{\mathrm{ch}} \approx 160$ MeV)
plays an essential role in describing yield ratios of hadrons. 
Thus, the dominant process of final hadron production
is supposed to change gradually from string fragmentation 
at the low-multiplicity limit to statistical hadronization at the high-multiplicity limit. 
At intermediate multiplicity, 
final hadron production would be a superposition of these two contributions.

The ``core--corona" picture has been adopted
to demonstrate the above idea 
\cite{Werner:2007bf,Aichelin:2008mi,Becattini:2008ya,Bozek:2005eu,Pierog:2013ria,Werner:2018yad}.
At the first contact of two nuclei, 
the number of collisions per nucleon should be different
depending on where they are located in the transverse plane being perpendicular to
the collision axis.
High-density regions just after the initial collision in which nucleons in colliding nuclei
suffer a lot of collisions are 
called ``core'' and assumed to be sources of the  QGP medium in local thermal and chemical equilibrium.
Low-density regions in which nucleons
suffer only a few collisions are 
referred to as ``corona'' and supposed to be likely to create hadrons 
via string fragmentation without formation of bulk medium. 
Since the ``core'' regions 
become dominant production mechanism with increasing multiplicity, 
the continuous change from string fragmentation  
to statistical hadronization with increasing multiplicity
could naturally be explained by the ``core--corona" picture.

In this paper, we introduce the core--corona picture
into the dynamical initialization framework to obtain the unified description 
of hadron yield ratios from small colliding systems to heavy-ion collisions.
The dynamical initialization framework \cite{Okai:2017ofp,Shen:2017bsr,Akamatsu:2018olk,Kanakubo:2018vkl}
describes the dynamics of gradual forming 
of the QGP fluids phenomenologically.\footnote{
The dynamical initialization framework is essential
 in description of nuclear collisions at lower energies
since hadrons are gradually generated 
due to insufficient Lorentz-contraction of the colliding nuclei
 \cite{Shen:2017bsr,Akamatsu:2018olk}.
}
Under this framework, the initial condition of the QGP fluids is obtained via energy-momentum deposition from the partons, strings, or fields 
generated just after the collision. 
As an extension of the dynamical initialization
description for generation of QGP fluids from initially produced partons
\cite{Okai:2017ofp,Shen:2017bsr,Akamatsu:2018olk,Kanakubo:2018vkl}, 
we introduced the dependence 
on initial parton densities in the dynamical initialization  \cite{Kanakubo:2018vkl},
which we call the ``dynamical core--corona initialization (DCCI) model'' in this paper. 
Given the various definitions of the core and the corona in the literature
\cite{Werner:2007bf,Aichelin:2008mi,Becattini:2008ya,Bozek:2005eu,Pierog:2013ria,Werner:2018yad},
we  here define them explicitly as follows:
The core represents the fluids under local thermal and chemical equilibrium, 
while the corona represents the system composed of nonequilibrated partons
traversing the fluids or the vacuum.
In the DCCI model,
the continuous separation
between the core and the corona is attributed to
the spatial density of initially produced partons.
Thus, it should be emphasized that 
we do not introduce the threshold to separate the core from the corona explicitly,
unlike the model in 
Refs.~\cite{Pierog:2013ria,Werner:2018yad}.
In the DCCI model, the nonequilibrated partons can coexist with the fluids
at some space-time point.

This paper is organized as follows:
In Sec.~\ref{sec:model}, we give an overview of
 the dynamical initialization framework
and its extension to the DCCI model.
In Sec.~\ref{sec:results}, 
we first investigate how much the energy is
fluidized at midrapidity when we reproduce the multiplicity dependence of particle ratio
from experimental data using the DCCI model.
Then we extract the fraction of fluidized energy as a function of multiplicity from experimental
data in both small and large colliding systems at LHC energies.
After that we discuss a universal behavior of the hadron yield ratios
as functions of multiplicity.
Finally Sec.~\ref{sec:summary} is devoted to summary of this paper.
  
Throughout this paper, 
we use the natural units, $\hbar = c = k_{B} =1$, and the Minkowski metric,
$g_{\mu \nu} = \mathrm{diag}(1, -1, -1, -1)$.

\section{Model}
\label{sec:model}
Before going into detailed explanations, 
we briefly give an outline of the DCCI model employed in this paper.
Initial partons 
are obtained
with the Monte Carlo event generator \pythia \ \cite{Sjostrand:2007gs}
and start traveling in the vacuum at their formation time. 
QGP fluids are dynamically generated via
energy-momentum deposition from these partons 
by solving the relativistic hydrodynamic equation with source terms. 
Here the four-momentum deposition per unit time 
is parametrized to be proportional to the initially produced parton density 
so that we take into account
the core--corona picture:
Partons in regions with larger density of surrounding partons
are more likely to deposit their four-momentum and generate QGP fluids,
 while those in dilute regions 
tend to traverse, keeping their four-momentum that they have just after the collision of nuclei.
In this way, separation of initially produced partons
into the QGP fluids (the core) and the non-equilibrated 
partons (the corona) is settled as a consequence of dynamical four-momentum deposition of
each parton. 
After the dynamical initialization with the core--corona picture,
we continue hydrodynamic evolution for fluids
until their temperature drops to the decoupling temperature.
Fluids are particlized into hadrons at the decoupling hypersurface 
via the Cooper--Frye formula \cite{Cooper:1974mv}.
On the other hand, surviving partons are hadronized 
under string fragmentation in \pythia. 
Therefore, the final hadron yield in this model is a summation of 
the contribution from particlization of the medium under local thermal and
chemical equilibrium
and that from string fragmentation.

\subsection{Hydrodynamic equations with source terms}
\label{subsec:eom}

In conventional hydrodynamic simulations,
the equation of continuity for energy and momentum,
\begin{equation}
\label{eq:equation-of-continuity}
\partial_\mu T^{\mu \nu} \!\left(x\right)= 0,
\end{equation}
is solved, where
$T^{\mu \nu}$ is the energy-momentum tensor of the fluid 
assuming that energy-momentum conservation is 
satisfied within the fluid. 
However, 
during the formation of the locally equilibrated medium in high-energy nuclear collisions,
there exist 
incoming energy and momentum for the formation of the medium
from non-equilibrated subsystems.
In our DCCI model, 
the initially produced partons 
deposit their energy and momentum
for the formation of the medium fluid 
and the energy-momentum conservation is imposed for 
the total system which is a composite one of the medium fluids and the traversing partons,
\begin{equation}
\label{eq:equation-of-continuity-fluids+partons}
\partial_\mu T_{\rm tot}^{\mu \nu} \!\left(x\right)= 
\partial_\mu \left(T_{\rm fluid}^{\mu \nu}(x) + T_{\rm parton}^{\mu \nu}(x) \right) = 0.
\end{equation}
Here we assume that energy and momentum deposited by partons are equilibrated instantaneously.
Equation (\ref{eq:equation-of-continuity-fluids+partons}) can be written in the form of
the relativistic hydrodynamic equation with source term,
\begin{equation}
\label{eq:qgp+jet}
\partial_\mu T_{\rm fluid}^{\mu \nu} \!\left(x\right)= J^\nu\!\!\left(x\right),
\end{equation}
if the source term is defined as 
\begin{equation}
\label{eq:definition-source-term}
J^{\nu} = - \partial_\mu T_{\rm parton}^{\mu \nu}.
\end{equation}
The hydrodynamic equation with source term~(\ref{eq:qgp+jet}) is also commonly used 
in the simulations  to discuss the physics of jet quenching and its effects on the medium
responses \cite{Tachibana:2014lja,Tachibana:2015qxa,Tachibana:2017syd,Chen:2017zte,Chang:2019sae,Tachibana:2020mtb}. 
If we assume ideal fluids,
the energy-momentum tensor of QGP fluids is decomposed as
\begin{equation}
  \label{eq:EMtensor}
T_{\rm fluid}^{\mu \nu}=(e+P)u^{\mu}u^{\nu} - Pg^{\mu \nu}.
\end{equation}
Here, $e$, $P$, and $u^{\mu}$ are energy density, hydrostatic pressure, 
and velocity of the fluid, respectively.
In order to close the equations of motion,
we need the equation of state (EoS) which has a form of
hydrostatic pressure as a function of energy density,
$P=P(e)$. In this study, we adopt the EoS 
from (2+1)-flavor lattice QCD calculations \cite{Borsanyi:2013bia}. 
In other words, $u$, $d$, and $s$ quarks (and their anti-quarks)
and gluons are under 
thermal and chemical equilibrium in the QGP fluids.

In this study, we do not solve the equation 
for the conservation of charges, such as baryon number, strangeness and electric charges,
since matter generated at RHIC and LHC energies is 
supposed to be almost baryon free, strangeness neutral, and charge neutral around midrapidity.

We also note that, 
although 
Eqs.~(\ref{eq:qgp+jet}) and 
(\ref{eq:EMtensor})
 are denoted in Cartesian coordinates
to avoid complex notations,
the actual calculation is performed in 
Milne coordinates.
In Milne coordinates, the time axis is represented with proper time 
$\tau = \sqrt{t^2 -z^2}$. 
The other spatial coordinates are $x$ and $y$, which are the transverse coordinates 
perpendicular to the collision axis,
and $\eta_{s} = (1/2)\ln[(t+z)/(t-z)]$, which is space-time rapidity.

\subsection{Dynamical initialization of QGP fluids}
\label{subsec:init}

In this study, the phase space distribution of initially produced partons 
is assumed to be
\begin{eqnarray}
\label{eq:parton-distibuiton-in-phasespace}
&&f_{\rm{parton}}(\bm{x},\bm{p}; t)d^3xd^3p \nonumber \\ 
& = & \sum_{i}  G(\bm{x}-\bm{x}_i(t)) \delta^{(3)}(\bm{p}-\bm{p}_i(t))d^3xd^3p,\\
&&G(\bm{x}-\bm{x}_i(t))d^3x =  \frac{1}{\sqrt{(2 \pi \sigma^{2})^3}}e^{-\frac{\left(\bm{x}-\bm{x}_{i}(t)\right)^{2}}{2 \sigma^{2}}} d^3x.
\label{eq:Gaussian_in_coordinate_space}
\end{eqnarray}
Here $\sigma$ is a width of Gaussian function in coordinate space. 
The Gaussian is introduced to give the scale of the region 
which is supposed to be involved in 
the interaction by a parton in the model. 
The trajectory of a parton is assumed to be eikonal, and 
rapidity of a parton is constant during the dynamical initialization,
\textit{i.e.},
 $y_i=\eta_{s,i}={\rm constant} $ where index $i$ represents the $i$th parton.
Under this assumption, the position of the $i$th parton is defined as
\begin{equation}
\label{eq:trajectory}
  \bm{x}_i(t)=\frac{{\bm{p}_i}}{p^0_i}(t-t_{{\rm form},i})+\bm{x}_{{\rm ini},i}, 
\end{equation}
where $\left(p^{0}_i, \bm{p}_i\right)$  
is the four-momentum of the $i$th parton.
Here, $\bm{x}_{{\rm ini},i}$ is the position 
at the formation time of the $i$th  parton, $t_{{\rm form},i}$.
Under these assumptions, we derive the explicit form of Eq.~(\ref{eq:definition-source-term}).
By putting the phase space distribution in Eq.~(\ref{eq:parton-distibuiton-in-phasespace})
into the kinetic definition of the energy-momentum tensor,
the source term (\ref{eq:definition-source-term}) becomes
\begin{eqnarray}
\label{eq:derivation-of-sourceterm-in-text}
J^\mu\left(x\right)
&=& - \partial_\mu \Tmunu_{\rm{parton}} \nonumber \\
&=& - \!\sum_{i} \!\int \!\!d^{3}p \frac{p^{\mu} p^{\nu}}{p^{0}} \roundmu f_{\rm{parton}}(\bm{x},\bm{p}; t)\nonumber \\
&=& -\sum_{i}  \frac{dp_i^\mu \left(t\right)}{dt} G\left(\bm{x}-\bm{x}_i(t)\right).
\end{eqnarray}
For details of this derivation, see Appendix \ref{sec:Derivation_of_fluidization_rate}.

We call ${dp_i^\mu}/{dt}$ in Eq.~(\ref{eq:derivation-of-sourceterm-in-text}) the ``fluidization rate",
which is the rate of four-momentum deposition for 
the $i$th parton.
Since the index $i$ represents each  parton produced in each event,
the summation is taken for all the initial partons in an event. 

We assume that QGP fluids are generated by the initial partons
during the proper time period
from the formation time $\tau=\tau_{00} (=0.1 \ \rm{fm})$ to initial time of 
fluids $\tau=\tau_0 (=0.6 \ \rm{fm})$.
In this paper, the common constant value of
$\tau_{00}$ is assumed for all partons,
$\tau_{{\rm form},i}=\sqrt{t_{{\rm form},i}^2 - z_{{\rm form},i}^2 } \equiv \tau_{00}$.
We start with the vanishing energy-momentum tensor of QGP fluids,
$T_{\rm fluid}^{\mu \nu}\left(\tau=\tau_{00}\right)=0$, 
and solve Eq.~(\ref{eq:qgp+jet}) from $\tau=\tau_{00}$ to $\tau=\tau_{0}$. 
Here we note that 
energy-momentum conservation is satisfied 
among the  QGP fluids and the traversing partons
through the dynamical initialization.
 
As we mentioned in the previous subsection, our actual simulations
are performed in (3+1)-dimensional Milne coordinates.
Thus the explicit form of the Gaussian distribution function 
in Eq.~(\ref{eq:derivation-of-sourceterm-in-text}) is replaced as
\begin{eqnarray}
\label{eq:gaussian}
&&G\left(\bm{x}-\bm{x}_i(t)\right)  d^{3}{x}\nonumber \\
& \rightarrow & \frac{1}{2\pi\sigma_\perp^2}\exp\left[-\frac{\left({\bm{x}}_\perp-{\bm{x}}_{\perp,i} (\tau)\right)^2}{2\sigma_\perp^2} \right]\nonumber \\
& \times & \frac{1}{\sqrt{2\pi \tau^2 \sigma_{\eta_{s}}^2}}\exp\left[-\frac{\left(\eta_{s}-\eta_{s,i}(\tau)\right)^2}{2\sigma_{\eta_{s}}^2} \right] \tau d\eta_{s} d^2x_\perp,
\end{eqnarray}
where ${\bm{x}}_{\perp,i}(\tau)$ and $\eta_{s,i}(\tau)$
are the transverse coordinates and the space-time rapidity of the $i$th parton.
In this study, we adopt $\sigma_\perp=0.5$ fm and $\sigma_{\eta_{\rm s}}=0.5$
for transverse and longitudinal widths of the Gaussian function, respectively.

\subsection{Energy-momentum deposition rate with the core--corona picture}
We introduce the core--corona picture as an extension of the 
dynamical initialization framework \cite{Kanakubo:2018vkl}.
To introduce the core--corona picture,
we parametrize the fluidization rate as
\begin{equation}
\label{eq:hydrorate}
\frac{d p_i^\mu}{dt}\left(t\right) = -a_0  \frac{\rho_i(\bm{x}_i(t))}{{p_{T, i}}^2} p_i^\mu  \left(t\right), \\ 
\end{equation}
where $p_{T, i}$ and $p_i^\mu$ are the transverse momentum and 
the four-momentum of the $i$-th parton, respectively.
Here, $\rho_i$ is the spatial density of partons surrounding the $i$th parton.
It is defined as 
\begin{eqnarray}
\label{eq:densitydistribution}
\rho_i(\bm{x}_i(t))d^{3}x
& = & \rho_i(\bm{x} = \bm{x}_i(t))d^{3}x \nonumber \\
& = & 
\sum_{j\neq i} \left. G\left(\bm{x}-\bm{x}_j\left(t\right)\right) d^{3}x \right|_{\bm{x} = \bm{x}_i\left(t\right)}.
\end{eqnarray}
The dimensionless factor $a_0$ is a free parameter to 
control the intensity of
energy-momentum deposition.
The Gaussian function $G$
has the same form as 
the one 
introduced in Eq.~(\ref{eq:parton-distibuiton-in-phasespace}) 
with the form of Eq.~(\ref{eq:gaussian}) in Milne coordinates.

The parton density $\rho_i$ introduced in  Eq.~(\ref{eq:hydrorate})
is the key factor to capture the feature of the core--corona picture. 
In high parton density regions, 
a large fraction of the energy and momentum of the parton is deposited to 
create the QGP fluids through the source term in Eq.~(\ref{eq:qgp+jet}). 
On the other hand, fluids are not likely to be generated in low parton density regions. 
It should be emphasized here that 
the criterion for the formation of the medium fluid is not governed by 
a certain threshold for the parton density but 
by the configuration of initial partons and their dynamics. 
In this framework,
medium fluids and nonequilibrated partons can even coexist at the same space-time point.

The factor $p_{T}^{\enskip -2}$ in the fluidization rate 
accounts for 
the tendency that lower $p_{T}$ partons are more likely to 
deposit their energy and momentum to form the fluids.
We should note that the power of $-2$ comes from consideration of dimension in Eq.~(\ref{eq:hydrorate}).
Although one may use more complicated forms of
the fluidization rate to capture the equilibration processes, we leave this consideration for our future work.

As we discuss in the next subsection,
the event generator \pythia, which we employ for generation of initial partons,
provides us with the color
flows of partons to form strings.
During the dynamical core--corona initialization,
we trace color flows of the partons in every time step 
and calculate the invariant mass of each string.
As the partons lose their energy and momentum, the invariant mass of strings becomes smaller and, 
eventually, some strings cannot undergo string fragmentation due to the lack of their invariant masses. 
To avoid such a situation, we assign the threshold of invariant mass for each string.
If the invariant mass becomes below
the threshold to undergo the string fragmentation in \pythia,
we put all the energy and momentum of partons in that string into fluids.
In this paper, 
we use the threshold of invariant mass as $m_{\rm th}=m_1+m_2 + 1.0$
in units of GeV for strings which have a quark/antidiquark and antiquark/diquark at the each end point.
Here $m_1$ and $m_2$ are the constituent masses of the leading quark/antidiquark and antiquark/diquark.

\subsection{Generation of initial partons}
In our DCCI framework, the event generator \pythia \ 8.230  \cite{Sjostrand:2007gs}
is used to simulate the initial parton production.
In \pythia,
we switch on the option to obtain partonic vertices (\texttt{PartonVertex:setVertex=on})
and switch off the hadronization (\texttt{HadronLevel:all=off}) 
to obtain the phase space distribution of partons at $\tau=\tau_{00}$.
\pythia \ provides us with the vertices for parton production
through multi parton interactions, final state radiations, and initial state radiations
\cite{Bierlich:2014xba}
which can be used in Eq.~(\ref{eq:trajectory}).
Since the Milne coordinates are employed in the actual simulations,
we take the transverse coordinates of the $i$th parton,
$\bm{x}_{\perp \textrm{ini},i}$, directly from \pythia \ and 
assume ${\eta_{s}}_{\textrm{ini},i}=y_{\textrm{ini},i}$, where $y_{\textrm{ini},i}$ is rapidity of a parton generated with \pythia.
Here, it should be noted that 
this version of \pythia \ handles heavy-ion reactions 
at high-energies using the Angantyr model 
\cite{Bierlich:2016smv,Bierlich:2018xfw}.

\subsection{Particlization of QGP fluids}
The dynamics of the medium after $\tau=\tau_0$ is treated in the same way as 
that in conventional hydrodynamic simulations. 
In this paper, we do not consider the energy and momentum loss 
of traversing partons due to the 
parton-medium interaction after $\tau=\tau_0$
for simplicity, and we solve Eq.~(\ref{eq:qgp+jet}) without the source term
until the maximum temperature of fluids becomes lower than 
the fixed decoupling temperature $T_{\mathrm{dec}}=160$ MeV. 
The effects of medium response on anisotropic flow
in the dynamical initialization
can be found in Ref.~\cite{Okai:2017ofp}.

We use the Cooper--Frye formula \cite{Cooper:1974mv}
to obtain spectra of hadrons emitted directly from the  decoupling hypersurface.
The yields from the medium fluids are obtained by integrating the spectra 
of each hadron species $i$
\begin{eqnarray}
N_i
&=&
\frac{g_i}{(2\pi)^3} \int \frac{d^3 p}{p^0}
\int_{\Sigma}
\frac{p^{\mu} d\sigma_{\mu}(x)}{\exp\left[ {p^{\mu}u_{\mu}\left(x\right)}/{T_{\mathrm{dec}}}\right]\mp_\mathrm{BF}1},
\label{eq:C-F}
\end{eqnarray}
where $g_i$ is the degeneracy, 
$\mp_\mathrm{BF}$ corresponds to Bose or Fermi distributions, 
$\Sigma$ is the decoupling hypersurface at $T=T_{\mathrm{dec}}$,
and $d\sigma_\mu$ is the normal vector of its hypersurface.
Since we assume baryon-free matter in this paper,
the chemical potential of baryon number does not appear in  Eq.~(\ref{eq:C-F}).
Here we should note that some fluid elements are already lower than the decoupling
temperature at $\tau = \tau_0$.
For such fluid elements,
we assume that the particlization is performed at the initial time of fluid $\tau=\tau_{0}$
with the corresponding temperature.

For the consideration of feed down from 
resonance decays, 
we correct the direct yields based on statistical model 
calculations \cite{Andronic:2017pug}.
We estimate the ratio of
the total yields to the contribution from directly produced 
hadrons $c_i$ from Fig.~2 of Ref.~\cite{Andronic:2017pug},
and multiply the ratio $c_i$ with the direct yield obtained from Eq.~(\ref{eq:C-F}).
Here we use the ratio factors,
$c_{\pi}=3.2$, $c_p=3.0$, $c_{\Lambda}=4.7$, $c_{\Xi}=1.7$, and $c_{\phi}=1.0$, 
 to obtain the total yields of these hadrons.

\subsection{String fragmentation of traversing partons}
We assume that the partons surviving after the dynamical core--corona initialization
form color singlet strings and are hadronized under the string fragmentation.
We push back the surviving partons into \pythia \ with 
their energy and momentum at $\tau=\tau_0$ (note that no parton energy loss
happens after $\tau=\tau_0$ in this paper)
and perform hadronization with the option \texttt{forceHadronLevel()} 
to get the final hadronic spectra.
Here we correct the energy of parton to 
be mass-on-shell using the momentum at $\tau=\tau_0$ and rest mass
for the execution of the string fragmentation,
since most partons are mass-off-shell
due to the four-momentum deposition of Eq.~(\ref{eq:hydrorate}).
We checked that the violation of energy 
conservation due to this correction is small enough to be ignored.
It should also be noted that 
we switch off weak decays of strange baryons which 
are stable against strong decays 
(except $\Sigma^0 \rightarrow \Lambda+\gamma$)
in \pythia \ .

\section{Results}
\label{sec:results}
The parton density distribution $\rho_i$ in the fluidization rate per parton (\ref{eq:hydrorate})
governs the separation of the core and the corona.
In the dynamical core--corona initialization, partons traversing the high-density region
tend to deposit their four-momentum
and to generate the fluids (the core).
In contrast,
partons traversing the low-density region
tend to stay as surviving partons (the corona).
As a result of the dynamical core--corona initialization,
the final hadron yields contain
the contributions from those two components.
The fraction of each contribution is
sensitive to distribution of the parton density from event to event.

In the following, the multiplicity
$\langle dN_{\mathrm{ch}}/d\eta \rangle$
describes the number of charged particles per unit pseudorapidity
in $|\eta|<0.5$ calculated
with default settings
in \pythia\.
To obtain the hadron yield ratios,
we first calculate final hadron yields from the core and the corona separately
and  the final hadron yield in each multiplicity class
is the sum of them,
\begin{eqnarray}
\label{eq:final_yield}
 \left\langle {\frac{dN_i}{dy}} \right\rangle = \left\langle \frac{dN_i}{dy} \right\rangle_{\rm{core}} 
+ \left\langle\frac{dN_i}{dy} \right\rangle_{\rm{corona}}.  
\end{eqnarray}
The angle bracket means the event average in a multiplicity class.
Although the final yield should be obtained as a sum of
two contributions from the core and the corona on an event-by-event basis,
the number of events for hydrodynamic simulations
is smaller than that for parton generation and hadronization in \pythia \
in order to reduce the computational cost.
In the results of the hadron yield ratios,
error bars represent statistical ones of yields from string fragmentation in \pythia,
while the shaded band represents statistical errors from the number of events 
in hydrodynamic simulations.
Since central collisions are more weighted
and the corresponding weight factor is provided for each event
 in the heavy-ion mode in \pythia,
we consider it in our statistical analysis.
We also note that the hadron yields from the core  are 
calculated via Eq.~(\ref{eq:C-F}) at $y=0$ assuming approximate boost invariance.
On the other hand,
the hadron yields from the corona are obtained
by counting the hadrons in $|y| < 2.0$ to gain sufficient statistics.

\subsection{Parameter $a_0$ dependence on particle ratio}
\label{sec:a_0_dep}
The free parameter $a_0$ in Eq.~(\ref{eq:hydrorate})
controls intensity of the fluidization
of partons besides the $\rho_i$. 
First we check the $a_0$ dependence 
on the particle ratio.
We perform simulations of Pb+Pb collisions at $\sNN = 2.76$ TeV
with $a_0 = 10, 20,$ and $100$ in the DCCI model.

\begin{figure}[htbp]
\begin{center}
\includegraphics[bb=0 0 432 540,width=0.5\textwidth]{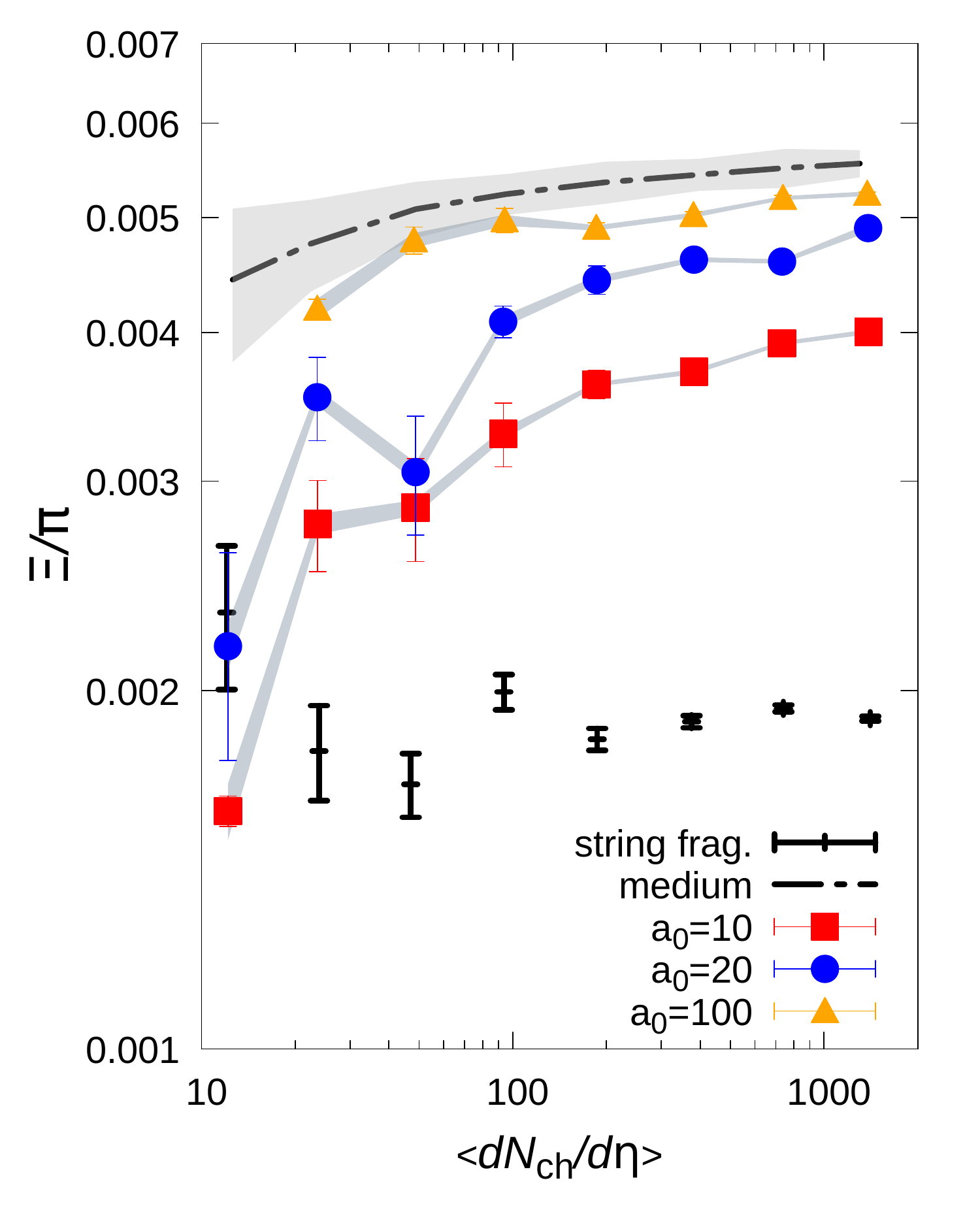}
\caption{(Color Online) Parameter $a_0$ dependence on hadron yield ratios of cascades to 
charged pions as functions of multiplicity in Pb+Pb collisions at $\sNN = 2.76$ TeV.
Results of $a_0 = 10$ (squares), $20$ (circles), and $100$ (triangles) are shown
for comparison.
The error bars of the hadron yield ratio
originate from the statistical errors of hadron yields from string 
fragmentations with \pythia,
while the shaded band 
comes from the statistical errors of yields from 
hydrodynamic simulations.
The hadron yield ratios only from the medium or string fragmentation are also shown as references.
}
\label{fig:a_0dependence}
\end{center}
\end{figure}
Figure \ref{fig:a_0dependence} shows 
the hadron yield ratio of cascades ($\Xi^{-} + \bar{\Xi}^{+}$) 
to charged pions 
($\pi^{+} + \pi^{-}$)
as a function of multiplicity, \textit{i.e.},
the event-averaged
charged hadron pseudorapidity density 
$\left<dN_{\mathrm{ch}}/d\eta\right>$ 
at midrapidity  $|\eta|<0.5$.
\footnote{
It should be noted that it is better to see the increasing behavior of the yield ratio of omega baryons since it shows more rapid enhancement.
However, we do not calculate them since it is statistically difficult to obtain enough yield of omega baryons within our framework. 
}
The hadron yield ratios in Pb+Pb collisions
at $\sqrt{s_{NN}}=2.76$ TeV
with three different $a_0$ parameters are shown as closed symbols.
We also show results with two limited cases: One with the case 
that all partons are fluidized and the other
with the default setting in heavy-ion mode in \pythia, which are referred as ``medium'' and ``string frag.''
in Fig.~\ref{fig:a_0dependence}.

As an overall tendency, the hadron yield ratio monotonically increases with multiplicity
and saturates in high-multiplicity events.
More specifically,  
results with $a_0 = 10$  and $20$ reflect
the value  from the corona $N_{\Xi}/N_{\pi} \approx 0.002$
at low-multiplicity and saturate towards the one from the core 
$N_{\Xi}/N_{\pi} \approx 0.005$
at high-multiplicity.
This is because the dominant contribution for the final yields changes 
from the corona to the core as multiplicity increases.
In addition to this, the hadron yield ratio with larger $a_0$ tends to saturate
at smaller multiplicity.
It can be naturally expected from the fact that 
the parameter $a_0$ controls 
the amount of four-momentum deposition from initial partons from Eq.~(\ref{eq:hydrorate}).
The results shown in Fig.~\ref{fig:a_0dependence} demonstrate this expectation.

We also note that the hadron yield ratios from both the fluids  and the string fragmentation 
do not seem to depend much on the multiplicity
since the hadron yield ratios from the core are determined from 
the decoupling temperature $T_{\mathrm{dec}}$ while the ones from the corona
are determined  from the string tension $\kappa$.

For a more quantitative view of how the parameter $a_0$ affects 
the hadron yield ratio as a function of multiplicity,
we fit the results shown in Fig.~\ref{fig:a_0dependence}
using the function 
\begin{equation}
f(x)=(F-S)\frac{x^n}{x^n + k^n}+S,
\label{eq:fitting}
\end{equation}
where the valuable $x$ denotes multiplicity.
Here we use $F=0.0055$ and $S=0.0018$ for
the hadron yield ratio of cascades to charged pions purely from  fluids 
and that from string fragmentation, respectively, assuming that the ratio from those 
components is constant as a function of multiplicity.\footnote{This function is motivated by the Hill equation which 
characterizes ligand bindings in biochemical reactions \cite{Hill1910}.}
The function in Eq.~(\ref{eq:fitting}) captures the 
behavior of the multiplicity dependence of hadron yield ratios, 
\begin{equation}
\lim_{x \to \infty} f(x) = F, \ 
\lim_{x \to 0+} f(x)= S.
\label{eq:limit-of-fitting-function}
\end{equation}
This fitting function has two fitting parameters, $k$ and $n$.
Using this function, we quantify the multiplicity, $x_{\mathrm{sat}}$, at which the hadron 
yield ratio reaches 90 \% of the difference between the
values obtained from the fluids and the one from string fragmentation,
namely, $f(x_{\mathrm{sat}})=S+0.9(F-S)$. 
Once the results are fitted, $x_{\mathrm{sat}}=k\sqrt[n]{9.0}$,
which is referred as saturation multiplicity, is obtained.
For details of the fittings, see Appendix \ref{sec:reduced_chisquare}.

\begin{figure}[htbp]
\begin{center}
\includegraphics[bb=0 0 648 504,width=0.5\textwidth]{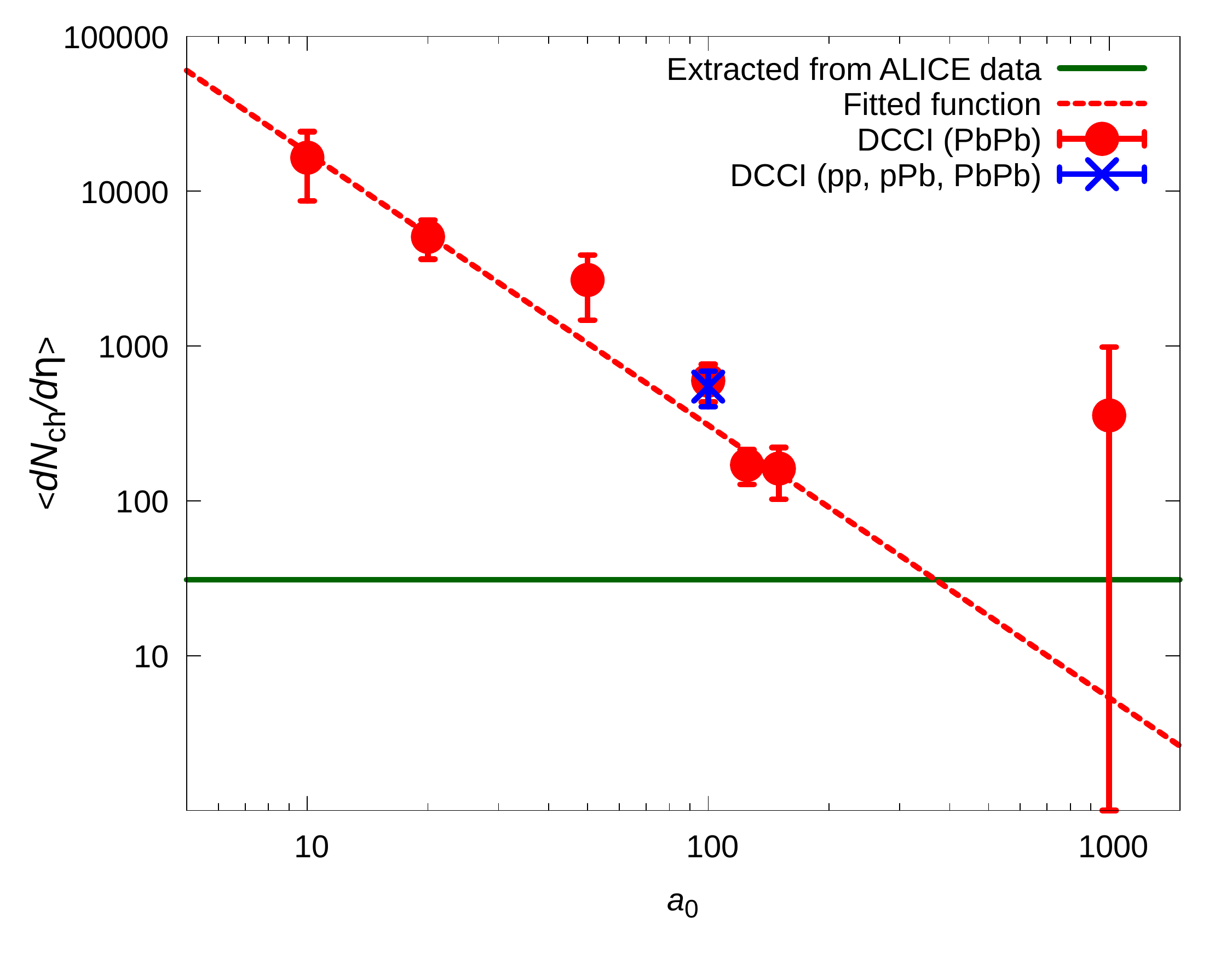}
\end{center}
\caption{(Color Online)
Parameter $a_0$ dependence of saturation multiplicity in Pb+Pb collisions (circles)
is compared with the result from simultaneous fitting for p+p, p+Pb, and Pb+Pb collisions (cross)
and that from fitting for the experimental data (horizontal line). 
The dashed line shows the fitting function for the saturation multiplicities (circles) fitted
by using a power function.
}
\label{fig:saturation-multiplicity}
\end{figure} 
Figure \ref{fig:saturation-multiplicity} shows the $a_0$ dependence of 
saturation multiplicity.
As expected from the results shown in Fig.~\ref{fig:a_0dependence},
the saturation multiplicity exhibits 
a clear monotonic decrease as a function of $a_0$.
We also simulate p+p collisions at $\sqrt{s}=7$ TeV and  p+Pb collisions
at  $\sNN=5.02$ TeV 
in addition to Pb+Pb collisions with $a_0 = 100$ in the DCCI model
and perform the function fittings of these results simultaneously.
Even if we include results from p+p and p+Pb collisions
in the fitting, the resultant 
saturation multiplicity is almost identical with
the one only from Pb+Pb collisions with the same $a_0$ value.
We also apply the global fitting for the ALICE experimental 
data in p+p, p+Pb, and Pb+Pb 
collisions \cite{ABELEV:2013zaa,Adam:2015vsf,ALICE:2017jyt}
and estimate its saturation multiplicity as $\left< dN_{\mathrm{ch}}/d\eta\right> \approx 31$,
shown as a solid line.
 
The value of the parameter $a_0$ which is most likely to reproduce 
the experimental data 
can be extracted by fitting 
the results obtained in the DCCI model
in Fig.~\ref{fig:saturation-multiplicity}.
Regarding the $a_0$ dependence of saturation multiplicity 
as a power function, we obtain
the optimized function $f(a_0)= a_0^{-1.8} \times 10^6$,
with reduced chi-square $\chi^2 \approx 1.2$.
This optimized function is shown as a dashed line in Fig.~\ref{fig:saturation-multiplicity}.
The value of $a_0$ at the intersection point of the solid and dashed lines in Fig.~\ref{fig:saturation-multiplicity}
enables us to reproduce the experimental data most reasonably within the DCCI model.
We estimate this value to be $a_0 \approx 368$.
Here, to obtain the optimized value for $a_0$, we use the central value of
the plots from the fitting for the experimental data in Fig.~\ref{fig:saturation-multiplicity}.

\subsection{Fraction of the fluidized energy }

We simulate p+p, p+Pb, and Pb+Pb collisions
with  $a_0=368$ in the DCCI model
to demonstrate how much energy and momentum just after the collisions 
are turned into the fluids.
We define the fraction of fluidized energy as
\begin{equation}
\label{eq:fraction-of-fluidized-energy}
R=\left.\frac{dE_{\mathrm{fluid}}/d\eta_{s}}{dE_{\mathrm{tot}}/d\eta_{s}}\right|_{\eta_{\rm s} = 0},
\end{equation}
which is  
the ratio of the energy density 
turned into the fluids to the total energy density at midrapidity ($\eta_{s} = 0$).
The fluidized energy can be obtained by integrating
the time component of source terms
from $\tau=\tau_{00}$ (formation time) to $\tau=\tau _0$ (hydrodynamic initial time)
in the transverse plane as
\begin{equation}
\label{eq:energy_fluid}
\left.\frac{dE_{\mathrm{fluid}}}{d\eta_{\rm s}}\right|_{\eta_{s} = 0}  =  \int_{\tau_{00}}^{\tau_{0}} d\tau \int   d^2x_\perp
 \tau J^\tau(\tau, x_\perp,  \eta_{s}=0).
\end{equation}
The total energy density
$dE_{\mathrm{tot}}/d\eta_{s}$ is 
calculated 
by taking the sum of all the initial partons' energy
at $\tau=\tau_{00}$.

\begin{figure}[htbp]
\begin{center}
\includegraphics[bb= 0 0 432 540,width=0.50\textwidth]{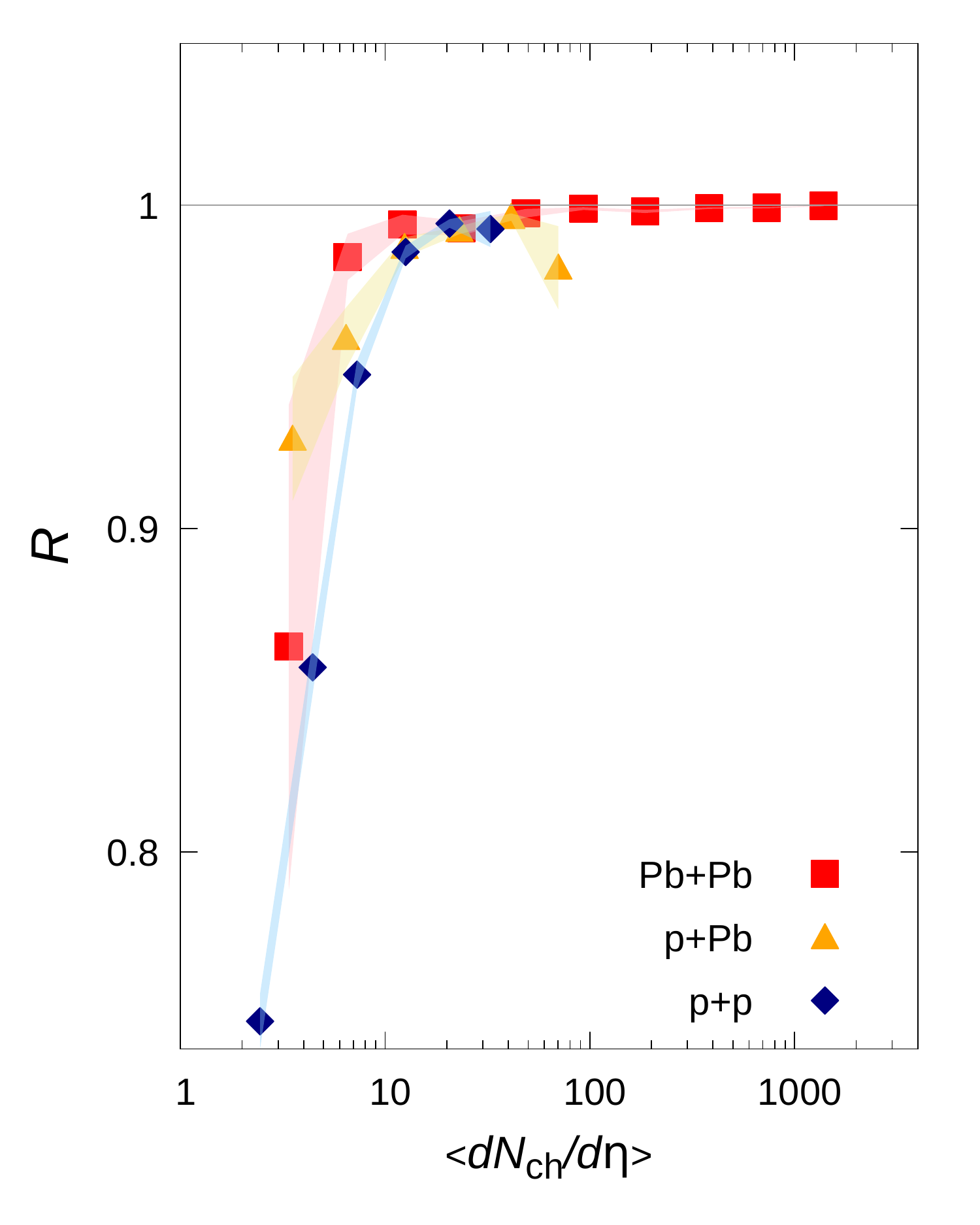}
\end{center}
\caption{(Color online)
Fraction of the fluidized energy to the total energy at
$\eta_{s} = 0$ 
as a function of multiplicity at midrapidity, $\left<dN_{\mathrm{ch}}/d\eta\right>$ ($|\eta|<0.5$),
in p+p (diamonds) collisions at $\sqrt{s}=7$ TeV,
p+Pb (triangles) collisions at $\sqrt{s_{NN}}= 5.02$ TeV, and  
Pb+Pb (squares)  collisions at $\sqrt{s_{NN}}= 2.76$ TeV. 
}
\label{fig:fluid-energy-fraction}
\end{figure} 

In Fig.~\ref{fig:fluid-energy-fraction},
the fraction of fluidized energy $R$ in p+p, p+Pb, and Pb+Pb collisions is shown
as a function of multiplicity.
The results exhibit the multiplicity scaling 
regardless of system sizes or collision energies.
$R$ increases monotonically 
and saturates around $\langle dN_{\mathrm{ch}}/d\eta\rangle_{|\eta|<0.5}\approx 20$. 
Note that hydrodynamic simulations are performed in the center-of-mass frame 
while multiplicities are calculated in
the laboratory frame. 
Therefore there is a rapidity shift $\Delta \eta_{s} = 0.47$ 
in p+Pb collisions at $\sNN = 5.02~\rm{TeV}$
between those frames.
The most remarkable thing is that the fraction of fluidized energy
$R$ is finite even at
the averaged multiplicity for nonsingle diffractive p+p
events, 
$\left<dN_{\mathrm{ch}}/d\eta\right> = 5.74\pm0.15$
\cite{Adam:2015gka}, which is generally considered to be too small to generate the QGP.
This means that, within our model calculations,
the contribution from the fluids to final hadron yields
is crucial to reproduce the yield ratio of cascades to pions 
in those multiplicity events.
Therefore this would be a strong indication of the partial QGP generation 
even in nonsingle diffractive p+p collisions at $\sqrt{s}=7$~TeV.

We should note that  the 
fraction of fluidized energy $R$ could be 
overestimated quantitatively
in the DCCI model.
Suppose
some hadronic strings are stretched 
between two surviving partons (possibly between a quark/antiquark and a diquark/antidiquark) 
each of which is in the projectile and in the target regions.
Although these strings do not contain partons
near midrapidity,
they can decay into hadrons in the 
midrapidity region.
We find $\approx$10\% of the yield at midrapidity originates
from these strings even when all partons are fluidized in the midrapidity region, \textit{i.e.},
$R \approx 1$.
It is expected that such strings would melt
when they penetrate the deconfined matter
and do not contribute to the final hadron yield at midrapidity.
This issue is to be resolved in future work.

\subsection{Hadron yield ratios}

\begin{figure*}[htbp]
\begin{center}
\includegraphics[bb=0 0 468 540,width=0.45\textwidth]{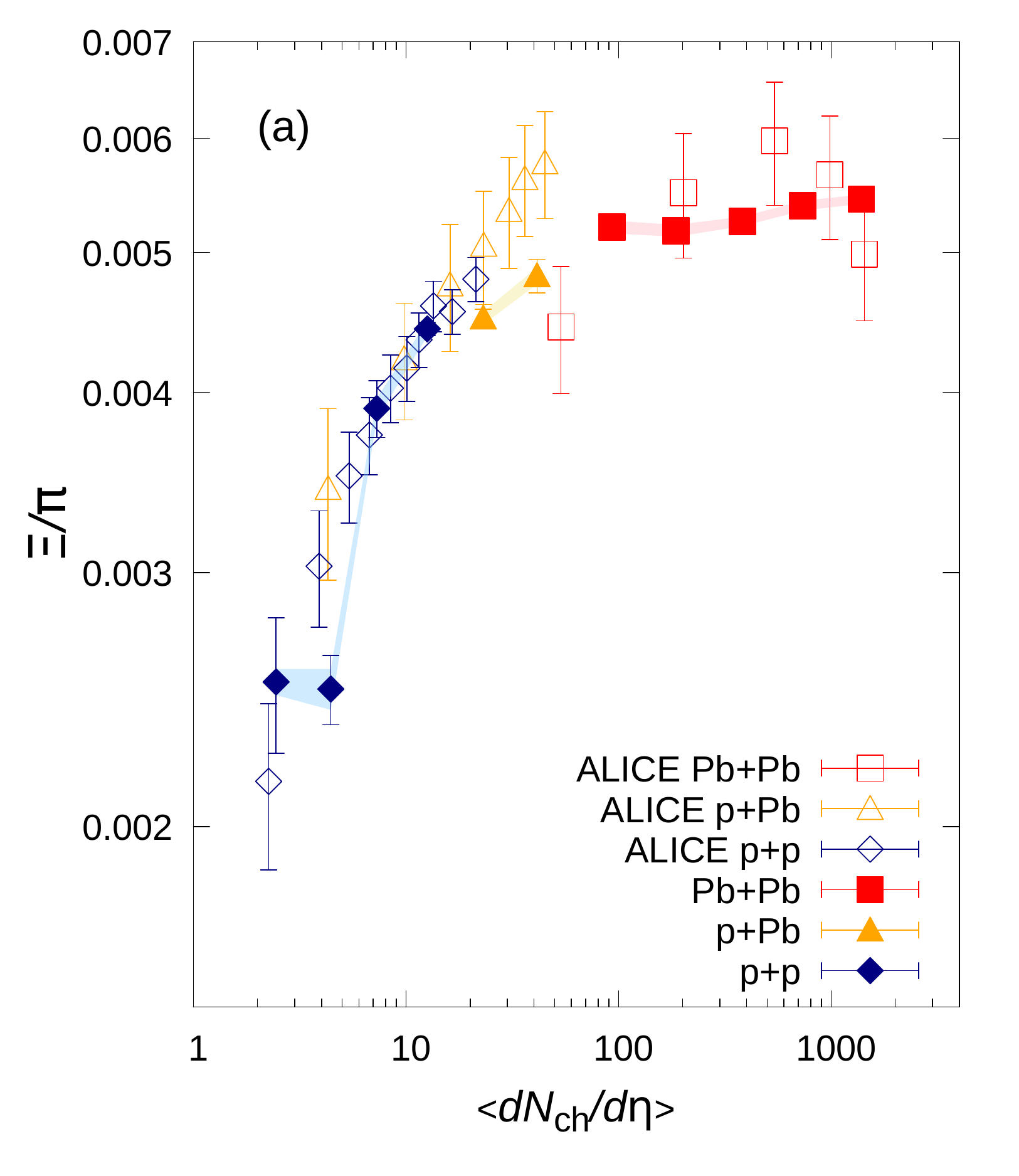}
\hspace{25pt}
\includegraphics[bb=0 0 468 540,width=0.45\textwidth]{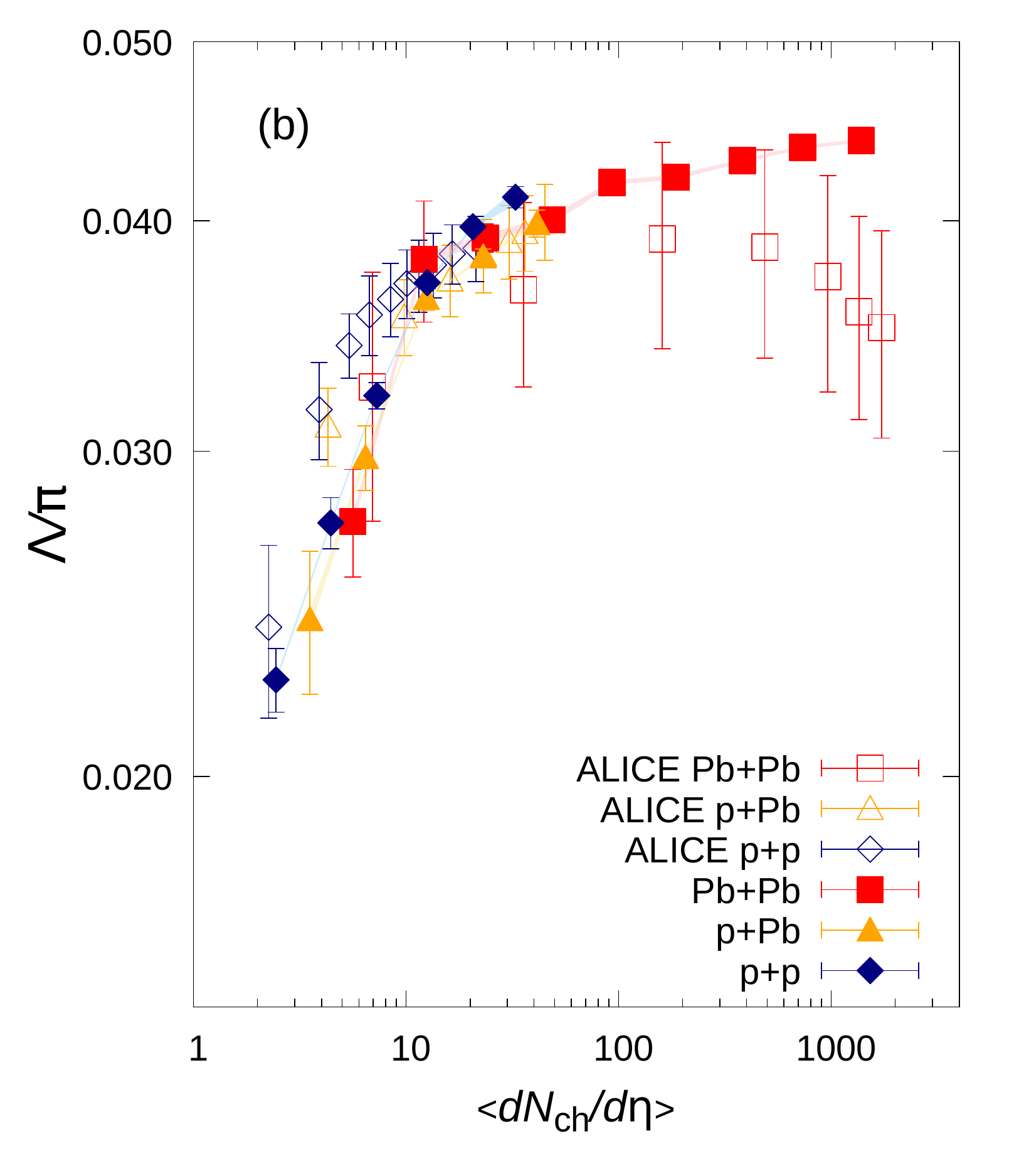}
\includegraphics[bb=0 0 468 540,width=0.45\textwidth]{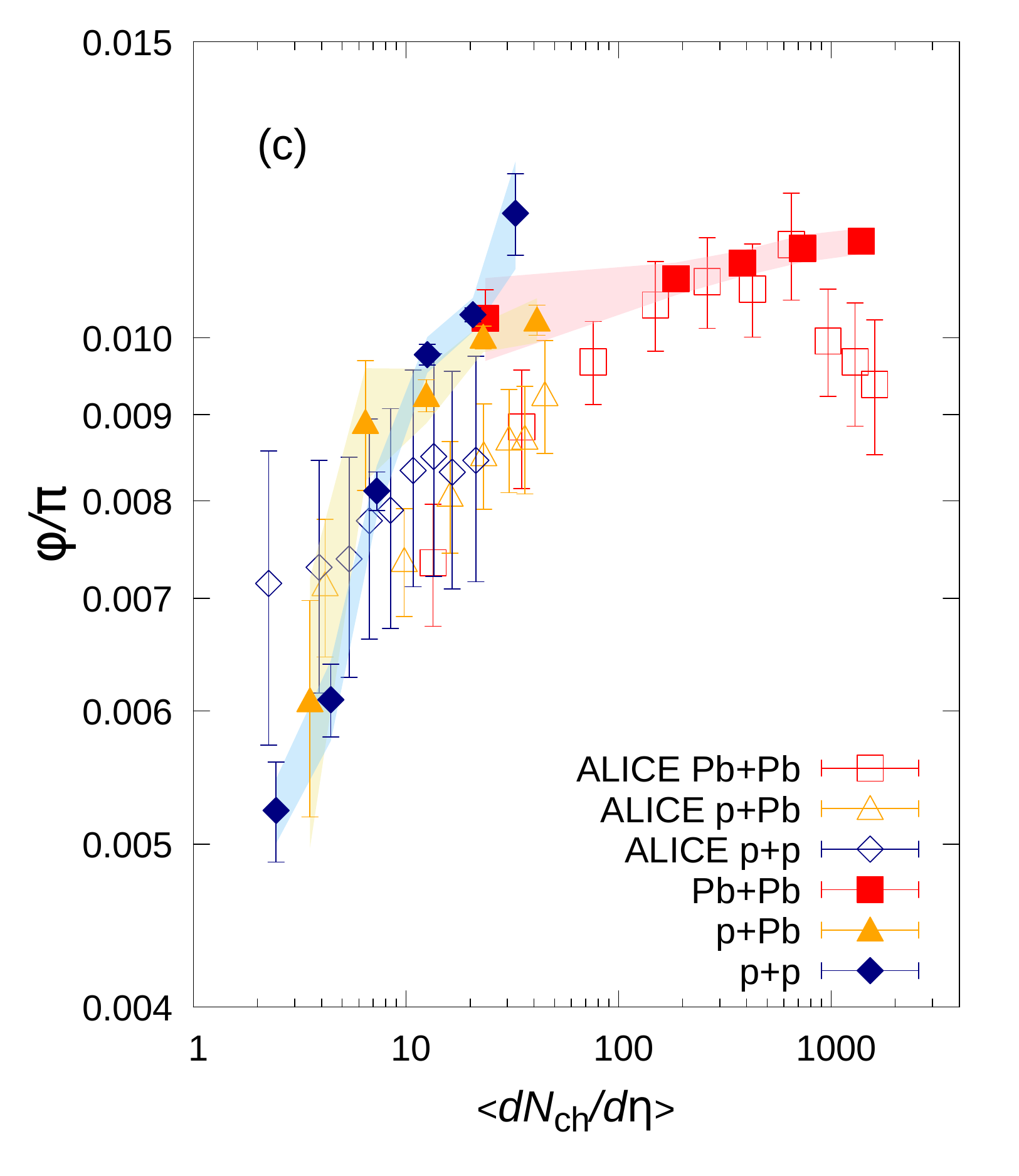}
\hspace{25pt}
\includegraphics[bb=0 0 468 540,width=0.45\textwidth]{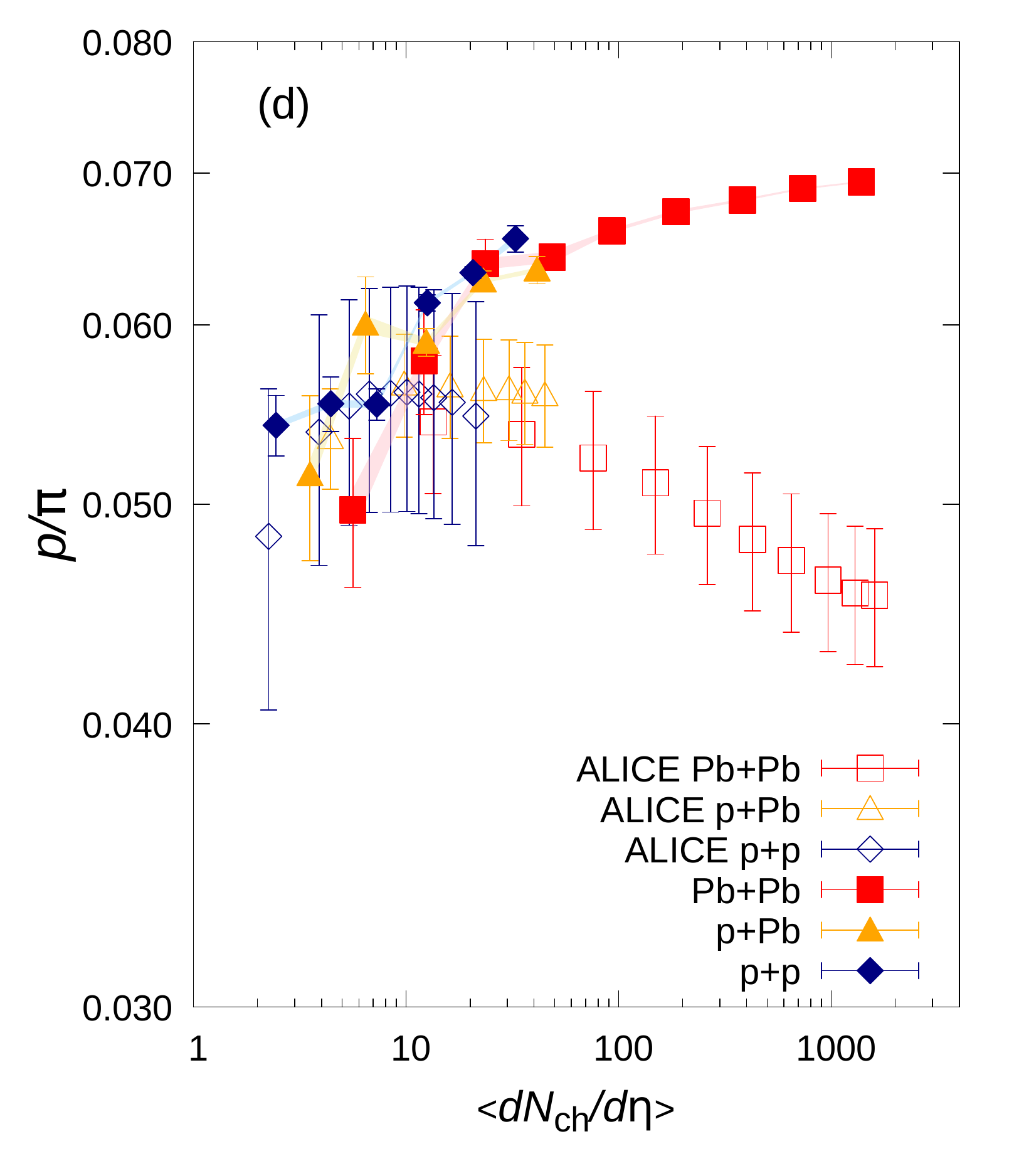}
\end{center}
\caption{(Color online)
Hadron yield ratios of (a) cascades ($\Xi^{-}$ and $\bar{\Xi}^{+}$),
(b) lambdas ($\Lambda$ and $\bar{\Lambda}$),  (c) phi mesons ($\phi$),
 and (d) protons ($p$ and $\bar{p}$) to charged pions ($\pi^-$ and $\pi^{+}$)
as functions of multiplicity at midrapidity, $\left<dN_{\mathrm{ch}}/d\eta\right>$,
in p+p (diamonds), p+Pb (triangles)
and Pb+Pb (squares) collisions at LHC energies.
The center-of-mass collision energies per nucleon pair are $\sqrt{s_{NN}}=7$ 
TeV (p+p), $5.02$ TeV (p+Pb), and $2.76$ TeV (Pb+Pb).
Results from the DCCI model (closed symbols) are
compared with the ALICE data (open symbols) 
\cite{Abelev:2014uua,Adam:2016bpr,Acharya:2018orn,Abelev:2013vea,Abelev:2013haa,Adam:2015vsf,ALICE:2017jyt}.
}
\label{fig:yield-ratios}
\end{figure*} 

We show the various hadron yield ratios 
in p+p, p+Pb, and Pb+Pb collisions as functions of multiplicity and 
compare them with ALICE data.
In Fig.~\ref{fig:yield-ratios},
the yield ratios as functions of
multiplicity  are shown for (a) cascades with $|S|=2$,
(b) lambdas with $|S|=1$, 
(c) phi mesons with $|S|=0$, 
and (d) protons and antiprotons. 
In the DCCI model, we use  $a_0=368$ extracted in Sec.~\ref{sec:a_0_dep}.
Since the parameter $a_0$ is fitted for experimental data  of cascades,
we reproduce the behavior of them well, as shown in Fig.~\ref{fig:yield-ratios}(a).
In addition, we also reproduce 
experimental data for lambdas and phi mesons reasonably well.
From Fig.~\ref{fig:fluid-energy-fraction}, it is clear that
the continuous change of hadron yield ratios results
from competition between contributions from the core and the corona.
Thus, continuous change of the dominant particle production mechanism
from the corona to the core enables the model to describe the experimental data.
These results also indicate that the DCCI model
describes the different tendencies of strangeness enhancement
among strange baryons which have different strangeness quantum numbers.

It is noted that 
the increasing behavior of the phi meson yield ratios
would not be
described by the canonical suppression model \cite{Acharya:2018orn}
since phi mesons are hadrons with hidden strangeness.
According to the fact that the models based on the core--corona picture, including our model,
show good agreement with the experimental data 
\cite{Werner:2018yad,Kanakubo:2018vkl,Kanakubo:2019ggp},
the core--corona picture turns out to be 
a more essential description of the multiplicity dependence of
hadron yield ratios compared to the canonical suppression model.

Regarding proton and antiproton yield ratios,
our results show continuous increase with multiplicity,
which is the same behavior as the other hadron yield ratios from the DCCI model.
However, the experimental data exhibit the opposite tendency:
The data decrease with multiplicity and deviate gradually from our results.
The value of the ratio of protons and antiprotons to charged pions 
from the DCCI model is $\approx 0.05$-0.06 in a few lowest-multiplicity bins,
which is almost the same value as the one that the corona gives.
On the other hand, the ratio is 
$\approx 0.07$ in the highest-multiplicity bin in this model,
which is almost consistent with the value obtained from the core. 
Thus the difference between our results and the
experimental data in high-multiplicity events indicates
a mechanism missing in the present model.
Since this deviation can be explained from baryon-antibaryon 
annihilation during hadronic evolution  \cite{Abelev:2013vea,Werner:2018yad},
we might resolve this issue by combining our model with hadronic cascade models
in the late stage.

\begin{figure}[htbp]
\begin{center}
\includegraphics[bb=0 0 360 540,width=0.5\textwidth]{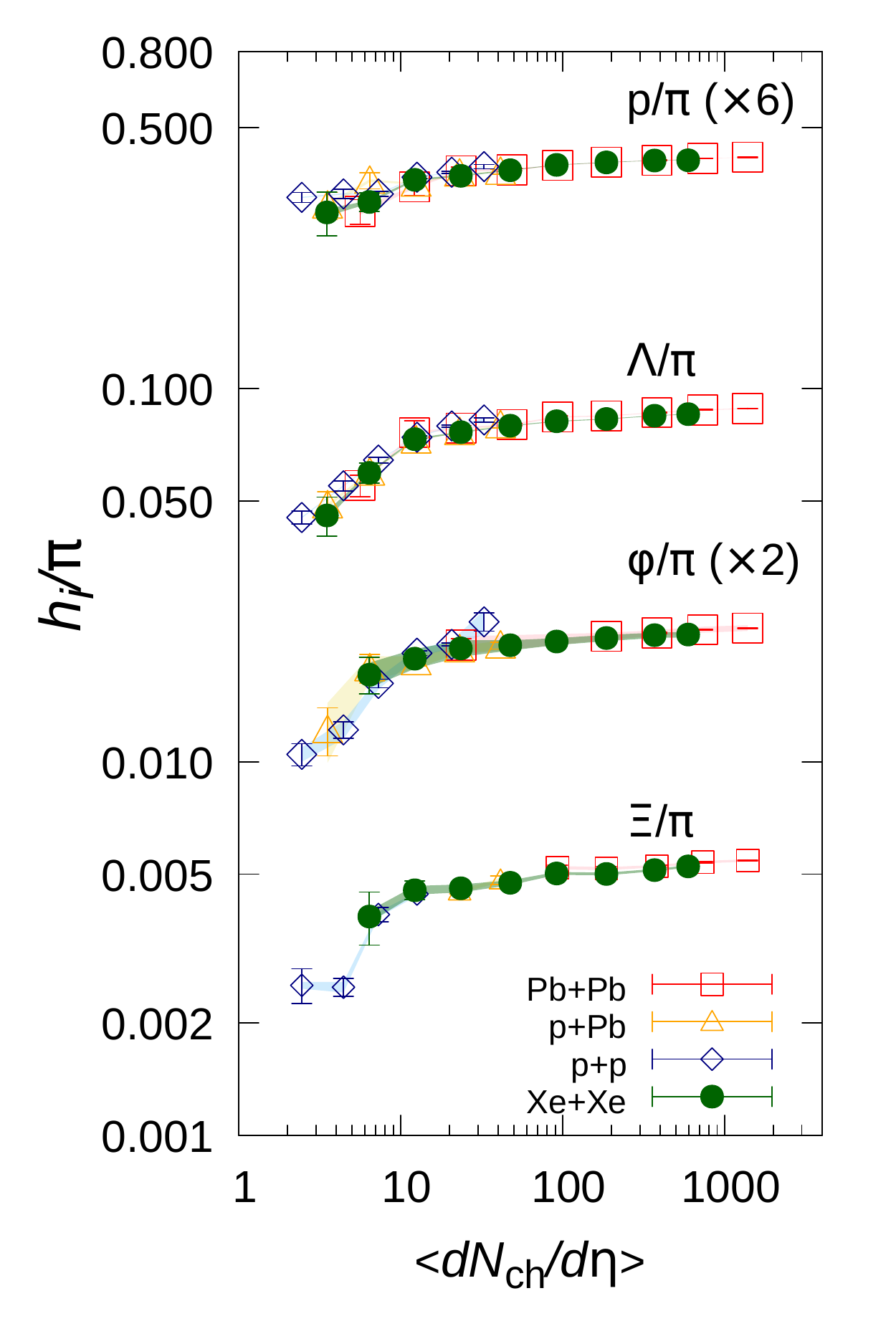}
\end{center}
\caption{(Color Online) Hadron yield ratios as functions of multiplicity 
in Xe+Xe collisions at $\sNN = 5.44$ TeV   (closed circles)
are compared with the ones in p+p (open diamonds),
p+Pb (open triangles), and Pb+Pb (open squares) collisions at LHC energies.}
\label{fig:XeXe}
\end{figure}

Motivated by the fact that 
hadron yield ratios scale with
multiplicity
regardless of system size or collision energy 
in the ALICE data \cite{ALICE:2017jyt}, 
we study whether or not the DCCI model gives the same tendency.
To see the system size independence at LHC energies,
the hadron yield ratios in Xe+Xe collisions at $\sNN=5.44$ TeV are 
compared with the ones in p+p, p+Pb, and Pb+Pb
collisions in Fig. ~\ref{fig:XeXe}.
The results in Xe+Xe collisions trace  the ones from the other collision systems,
which exhibit the same behavior as ones 
reported in Ref.~\cite{Albuquerque:2018kyy}.

\begin{figure*}[htbp]
\begin{center}
\includegraphics[bb=0 0 360 540,width=0.45\textwidth]{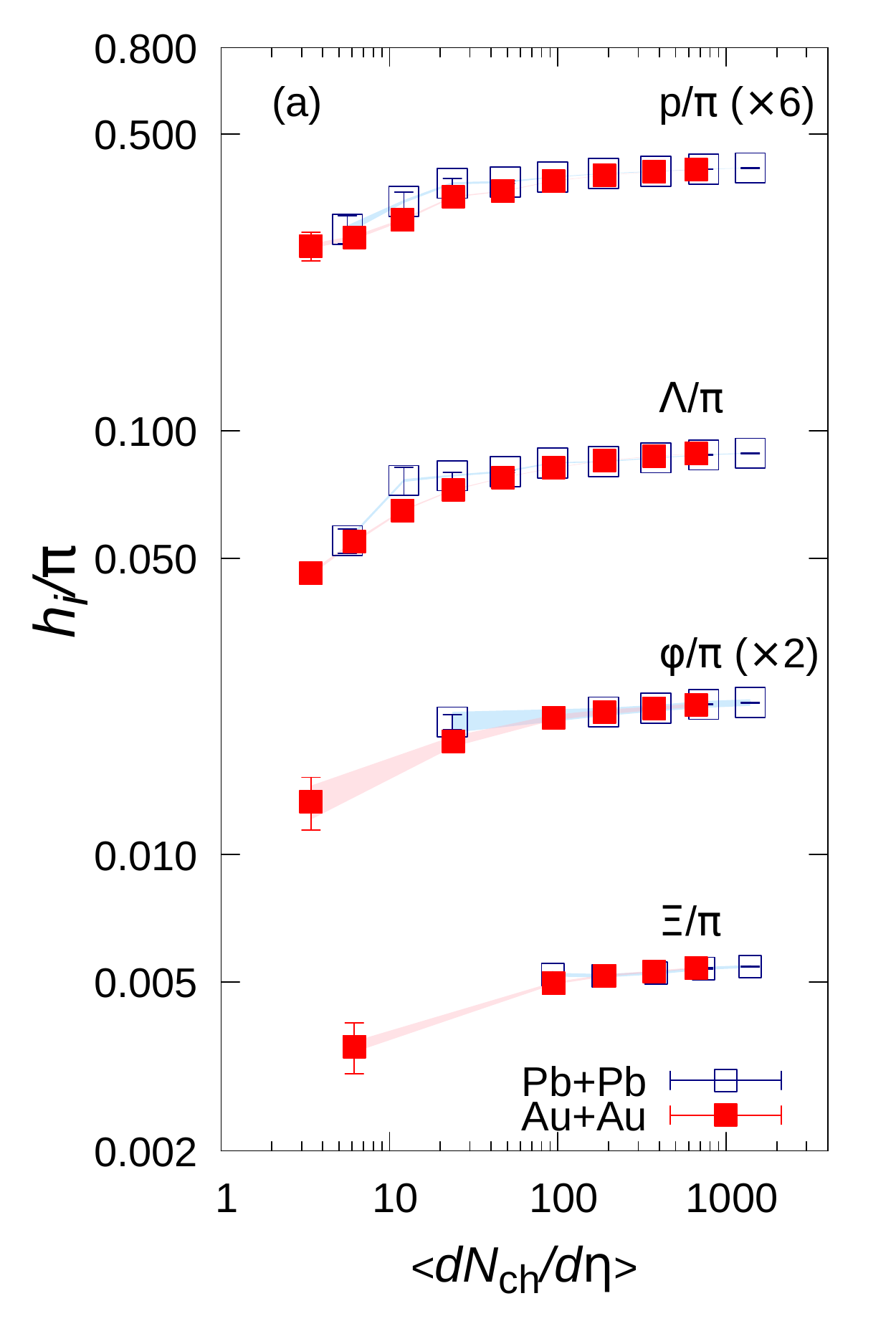}
\includegraphics[bb=0 0 360 540,width=0.45\textwidth]{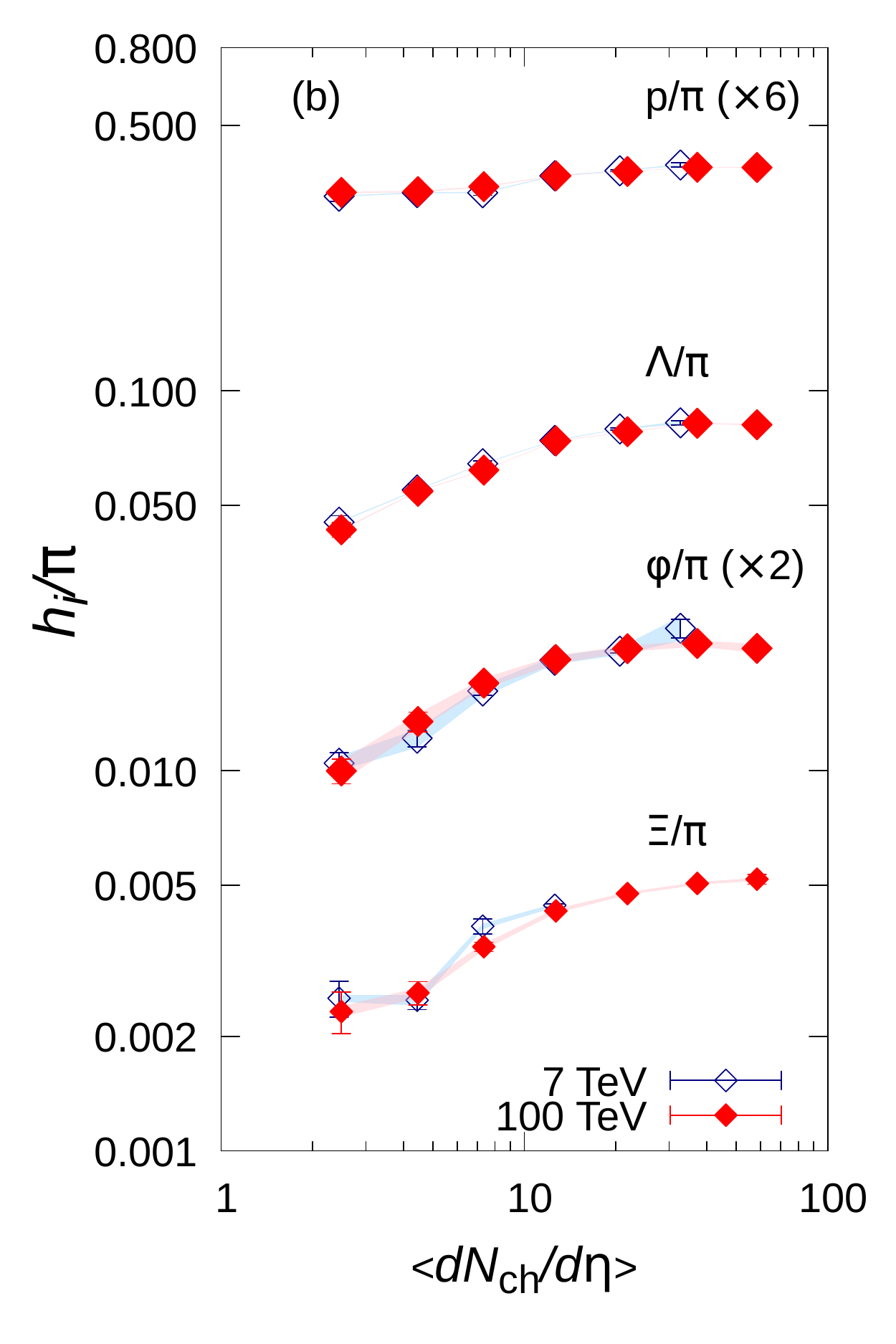}
\end{center}
\caption{(Color Online)
(a) Hadron yield ratio as a function of multiplicity in heavy-ion collisions at RHIC and LHC energies.
Results in Pb+Pb collisions at $\sNN = 2.76$ TeV 
(open symbols) are compared with the ones in Au+Au collisions at $\sNN = 200$ GeV (closed symbols).
(b) Hadron yield ratio as a function of multiplicity 
in p+p collisions at LHC and FCC energies.
Results at $\sqrt{s} = 7$ TeV (open symbols) are compared with
the ones  at $\sqrt{s} = 100$ TeV (closed symbols).
}
\label{fig:RHIC-FCC}
\end{figure*}

Figures~\ref{fig:RHIC-FCC}(a) and \ref{fig:RHIC-FCC}(b)
show collision energy (in)dependence of hadron yield ratios
as functions of multiplicity in nuclear and p+p collisions, respectively.
The results in Au+Au collisions at $\sNN=200$ GeV 
are compared with the ones in
 Pb+Pb collisions at $\sNN=2.76$ TeV in Fig.~\ref{fig:RHIC-FCC}(a).
On the other hand, a comparison between p+p collisions at
 $\sqrt{s}=7$ and $100$ TeV
is made in Fig.~\ref{fig:RHIC-FCC}(b).
It is noted that this extremely large collision energy of $\sqrt{s}=100$ TeV is
the same as the one planned at the Future Circular Collider (FCC) experiment.
The tendency is the same as that of the other results again: The hadron yield ratios 
exhibit scaling behavior as functions of multiplicity 
and there is almost no collision 
energy dependence regardless of a gap of
collision energy of one or two orders of magnitude for each comparison.

These results demonstrate absence of 
size and energy dependence for hadron yield ratios as functions of multiplicity.
Therefore it can be said that the multiplicity is 
one of the keys to control the fraction of energy converted into the QGP fluids in high-energy nuclear collisions. 

\section{Summary}
\label{sec:summary}

In this study, we analyzed multiplicity dependence of
the hadron yield ratios in various collision systems and in a wide range of collision energy,
based on the dynamical core--corona initialization framework,
and studied how much in fraction the equilibrated matter is formed in p+p, p+Pb, and Pb+Pb collisions.

Assuming that fluids are generated via 
four-momentum deposition of the initially produced partons,
we described the initial stage of forming the QGP fluids
with hydrodynamic equations with source terms.
We formulated the source term 
considering the spatial geometry of the initial partons
under the concept of the core--corona picture.
In this picture,
partons in the high-density region are likely to be components
of the QGP fluids, while those in the low-density region tend
to survive as partons traversing the QGP fluids or vacuum.
Using the dynamical core--corona initialization framework,
we performed numerical simulations of
nucleon--nucleon, nucleon--nucleus, and nucleus--nucleus 
collisions at RHIC, LHC, and FCC energies.
First we generated initial partons with \pythia \ 8.230, switching off the hadronization.
Next we performed the dynamical core--corona initialization  
from the formation time to the initial time of fluids.
QGP fluids are generated through the source term which is expressed as
the sum of four-momentum deposition of each parton generated initially.
The core and the corona are separated in accordance with density of initial partons
based on the core--corona picture.
The space-time evolution of QGP fluids (the core)
is described by (3+1)-dimensional ideal hydrodynamics.
The surviving partons (the corona) keep traversing the QGP fluids or vacuum.
Each system undergoes different hadronization processes.
In the conventional way, we converted the QGP fluids to hadrons via the Cooper--Frye formula
at the decoupling temperature.
On the other hand, the traversing partons form color singlet
strings and are hadronized via string fragmentation with \pythia.
Final hadron yields are composed of production of both the core and the corona.
Since each hadronization mechanism gives the proper value of particle ratios,
the ratios in the final hadron production reflect the competition between two contributions.
As a result of the dynamical core--corona initialization, the fraction of the core
increases monotonically with multiplicity, and hadron yield ratios eventually change from the value of the string fragmentation
to the one of statistical models.

We analyzed the multiplicity dependence of yield ratios of cascades to charged pions
with various values of $a_0$, which is a parameter to control the fraction of the core
in the fluidization rate.
Performing chi-square fitting for the yield ratios of  our results
 and the experimental data,
we found that $a_0=368$ would be the best value to describe the 
experimental data within our model.
Using the extracted $a_0$, we performed simulations 
of p+p, p+Pb, and Pb+Pb collisions at LHC energies
and quantified the fraction of the fluidized energy at midrapidity.
Our result showed that the fraction of fluidized energy
increases monotonically with the multiplicity.
We concluded that even at the multiplicity 
in nonsingle diffractive p+p events there exists hadron production from
the chemically equilibrated matter 
to reproduce the monotonically increasing behavior of multiplicity dependence of hadron yield ratios 
reported by the ALICE Collaboration.
We also analyzed
the multiplicity dependence of yield ratios of cascades, lambdas, phi mesons, and protons
with the extracted $a_0$ in various colliding systems and in a wide range of collision energy.
Except for protons, our model calculations showed reasonable 
agreement with the ALICE experimental data.
In the low-multiplicity events $\left<dN_{\rm{ch}}/d\eta\right> \approx 2$-$3$,
hadron yield ratios are almost identical with those from
string fragmentation, while in high-multiplicity events,
above $\left<dN_{\rm{ch}}/d\eta\right> \approx  300$,
hadron yield ratios are almost the same as those of statistical models.
We described the continuous increase of hadron yield ratios observed in the experimental data 
as a result of competition between the core and the corona.
Although our results describe the experimental data reasonably, we admit that their detailed characteristics are not reproduced. Nevertheless there is still room to introduce additional dynamics such as hadronic rescatterings and viscosity. We leave implementation of them as our future work.
In the dynamical core--corona initialization model, 
system size and collision energy dependences of the 
hadron yield ratio  were analyzed
by performing simulations of
p+p, Xe+Xe, and Au+Au collisions at $\sNN=100$ TeV, $5.44$ TeV, and $200$ GeV,
respectively.
We studied whether there is a system size dependence
of the hadron yield ratio as a function of multiplicity
by comparing the results of Xe+Xe collisions with the ones of other 
collision systems at LHC energies.
We also checked whether the collision energy dependence appears
when comparing results at RHIC energy  with the ones at LHC energy,
and results in p+p collisions at FCC energy with the ones at LHC energy.
Since both results show clear scaling with multiplicity,
there are no system size and collision energy dependences
within the dynamical core--corona initialization model.

In this work, we focused on the bulk property 
of the system created in high-energy nuclear collisions and 
saw that the dynamical core--corona initialization model
reasonably describes the
multiplicity dependence of strange hadron yield ratios from 
small to large colliding systems.
On the other hand, 
it is still not revealed to what extent
collectivity observed in small colliding systems
originates from the collective hydrodynamic flow of the QGP fluids.
Flow observables  in small colliding systems have been discussed frequently
 in the context of QGP formation theoretically and experimentally.
We point out here that 
the bulk properties such as hadron yield ratios also should be explained
as well as the flow observables within the same framework.
Conventionally viscous hydrodynamic models, which were successful in description of
flow observables in high-energy heavy-ion collisions, have been applied also to
small colliding systems. However, even if these hydrodynamic models 
reproduce flow data in small colliding systems,
they might not reproduce the bulk properties, in particular, hadron yield ratios.
In the light of our analysis and interpretation of hadron yield ratios data,
there should exist a certain contribution from nonequilibrated
systems in the final hadron production which would dilute collective flow signals to some extent.
Therefore in our future work we plan to study to what extent hydrodynamic flow signals
generated by the core
would be affected by the corona as a nonequilibrated system
within the dynamical core--corona initialization model.
In order to perform sophisticated analysis, we will introduce viscosities in fluids and
hadronic rescatterings in the late stage in the dynamical core--corona initialization model.

It is also worth studying the effect of dynamical core--corona initialization
on transverse dynamics in high-energy nuclear collisions.
We anticipate that high $p_{T}$ partons behave as the corona and traverse the core after 
the dynamical core--corona initialization.
Thus we separate hard from soft components naturally and dynamically.
This would be a starting point to investigate the modification of jet structure
in a hydro-based full event generator.
These will be discussed in our future publications.

\section*{Acknowledgments}
The work by Y.T. is supported in part by a special grant from 
the Office of the Vice President of Research at Wayne State University 
and in part by the National Science Foundation (NSF) 
within the framework of the JETSCAPE Collaboration under Grant No. ACI-1550300.
The work by T.H. is partly supported by JSPS KAKENHI Grant No. JP17H02900.\\

\appendix
\section{Derivation of fluidization rate}
\label{sec:Derivation_of_fluidization_rate}

The fluidization rate is obtained by substituting Eq.~(\ref{eq:parton-distibuiton-in-phasespace}) 
into the kinetic definition of energy-momentum tensor in Eq.~(\ref{eq:definition-source-term}).
The right hand side of  Eq.~(\ref{eq:definition-source-term}) without minus sign
becomes
\vspace{0.5cm}
\begin{eqnarray}
\label{eq:derivation-of-sourceterm}
  && \partial_\mu \Tmunu_{\rm{parton}} \nonumber \\
  &=& \sum_{i} \int d^{3}p \frac{p^{\mu} p^{\nu}}{p^{0}} \roundmu G(\bm{x}-\bm{x}_i(t)) \delta^{(3)}(\bm{p}-\bm{p}_i(t)) \nonumber \\
  &=& \sum_{i} \int d^{3}p \frac{p^{\mu} p^{\nu}}{p^{0}} \left[ \bigl( \roundmu G(\bm{x}-\bm{x}_i(t)) \bigr) \delta^{(3)}(\bm{p}-\bm{p}_i(t)) \right. \nonumber \\ 
    &&    \left.  + G(\bm{x}-\bm{x}_i(t)) \bigl( \roundmu \delta^{(3)}(\bm{p}-\bm{p}_i(t)) \bigr) \right]. \nonumber\\
\end{eqnarray}
\vspace{0.5cm}
The first term in Eq.~(\ref{eq:derivation-of-sourceterm}) vanishes as
\begin{eqnarray}
&& \sum_{i} \int \frac{d^{3} p}{p^{0}}  p^{\nu}\left[ \left(p^{0} \frac{\partial}{\partial t}+\bm{p} \cdot \nabla\right) \frac{1}{\sqrt{(2 \pi \sigma^{2})^3}} e^{-\frac{\left(\bm{x}-\bm{x}_{i}(t)\right)^{2}}{2 \sigma^{2}}} \right]\nonumber \\
&& \times  \delta^{(3)}(\bm{p}-\bm{p}_i(t)) \nonumber \\
&=&\sum_{i} \int \frac{d^{3} p}{p^{0}} p^{\nu} \left[ p^{0} \frac{{\bm{x}}-{\bm{x}}_i (t)}{\sigma^{2}} \cdot \frac{d \bm{x}_{i}}{dt}- \bm{p}_i \cdot  \frac{{\bm{x}}-{\bm{x}}_i (t)}{\sigma^{2}}   \right] 
  \nonumber \\
  &&   \times  \frac{1}{\sqrt{(2 \pi \sigma^{2})^3}} e^{-\frac{\left(\bm{x}-\bm{x}_{i}(t)\right)^{2}}{2 \sigma^{2}}} 
\delta^{(3)}(\bm{p}-\bm{p}_i(t)) \nonumber\\
&=&0. \nonumber
\end{eqnarray}
\vspace{0.5cm}
Here we use $p^0_i d\bm{x}_i/dt = \bm{p}_{i}$.
On the other hand, the second term in Eq.~(\ref{eq:derivation-of-sourceterm}) is
\begin{eqnarray}
&& \sum_{i} \int \frac{d^{3} p}{p^{0}} p^{\nu}G\left({\bm{x}}-{\bm{x}}_{i}\right)p^{\mu} \roundmu \delta^{(3)}(\bm{p}-\bm{p}_i(t)) \nonumber\\
&=&\sum_{i} \int \frac{d^{3} p}{p^{0}} p^{\nu}G\left({\bm{x}}-{\bm{x}}_{i}\right)p^{0} \frac{\partial}{\partial t}  \frac{1}{(2 \pi)^{3}} \int d^{3} x^{\prime} e^{i({\bm{p}}-{\bm{p}}_i(t)) \cdot \bm{x}^{\prime}} \nonumber\\
&=&-\frac{i}{(2 \pi)^{3}} \sum_{i} G({\bm{x}}-{\bm{x}}_i) \nonumber \\
& & \times \int \frac{d^{3} p}{p^{0}} p^{\nu} p^{0} 
 \int d^{3} x^{\prime} \frac{d {\bm{p}}_{i}}{dt} \cdot {\bm{x}}^{\prime} e^{i({\bm{p}}-{\bm{p}}_i (t)) \cdot \bm{x}^{\prime}}. \nonumber
\end{eqnarray}
\vspace{0.5cm}
\noindent Finally Eq.~(\ref{eq:derivation-of-sourceterm}) becomes
\begin{eqnarray}
&&\partial_\mu \Tmunu_{\rm{parton}} \nonumber \\
&=&-\frac{1}{(2 \pi)^{3}} \sum_{i} G({\bm{x}}-{\bm{x}}_i)\nonumber \\
& & \times \int \frac{d^{3} p}{p^{0}} p^{\nu} p^{0} \frac{d {\bm{p}}_{i}}{dt} \cdot \frac{\partial}{\partial {\bm{p}}} \int d^{3} x^{\prime} e^{i({\bm{p}}-{\bm{p}}_i) \cdot \bm{x}^{\prime}} \nonumber\\
&=& - \sum_{i} G({\bm{x}}-{\bm{x}}_i) \int \frac{d^{3} p}{p^{0}} p^{\nu} p^{0} \frac{d {\bm{p}}_{i}}{dt} \cdot \frac{\partial}{\partial {\bm{p}}} \delta^{(3)}(\bm{p}-\bm{p}_i) \nonumber\\
&=&\sum_{i} G({\bm{x}}-{\bm{x}}_i) \int {d^{3} p}  \frac{d {\bm{p}}_{i}}{dt} \cdot \frac{\partial p^{\nu} }{\partial {\bm{p}}} \delta^{(3)}(\bm{p}-\bm{p}_i) \nonumber\\
&=&\sum_{i} G({\bm{x}}-{\bm{x}}_i) \frac{d p_i^{\nu}}{dt}. \nonumber
\end{eqnarray}
Thus the source term (\ref{eq:definition-source-term}) is derived as
\begin{equation}
\label{eq:source_j}
J^\mu\left(x\right)
=  -\sum_{i}  \frac{dp_i^\mu \left(t\right)}{dt} G\left(\bm{x}-\bm{x}_i(t)\right),
\end{equation}
by assuming the phase space distribution as Eq.~(\ref{eq:parton-distibuiton-in-phasespace}).

\section{Reduced chi-square values of function fitting}
\label{sec:reduced_chisquare}
Table~\ref{table:xipi-data} shows the reduced chi-square values
of function fitting for yield ratios of cascades to pions as functions of multiplicity 
with various $a_0$ parameters shown in Fig.~\ref{fig:a_0dependence}.
Since  the statistical errors  from hydrodynamic simulations are relatively small, 
only the statistical errors due to string fragmentation are considered
in the chi-square fitting. 
We also show the reduced chi-square value from function fitting 
for the yield ratio of cascades  to pions in ALICE experimental data \cite{ALICE:2017jyt}.
\begin{table}[htb]
\begin{center}
  \begin{tabular}{ccccc} 
    \hline \hline 
    $a_0$ & Collision system & $k$ & $n$ & Reduced $\chi^2$ \\ 
    \hline
     10 & Pb+Pb &  577.785 & 0.656 &  35.8272 \\
     20 & Pb+Pb &  43.195 & 0.461 & 3.61187 \\    
     50 & Pb+Pb &  3.03 & 0.324 & 2.87592 \\
     100 & Pb+Pb &  4.985 & 0.459 & 2.37556 \\
     100 & p+p, p+Pb, and Pb+Pb & 8.595 & 0.529 & 11.2227 \\ 
     125 & Pb+Pb & 6.745  & 0.68 & 4.92507 \\
     150 & Pb+Pb & 1.16 & 0.445 & 21.4127 \\
     1000 & Pb+Pb & 0.095 & 0.267 & 4.87633 \\ 
     N/A & p+p, p+Pb, and Pb+Pb & 6.685 & 1.434 & 0.971169 \\ 
     \hline \hline
  \end{tabular}
\end{center}
\caption{Reduced chi-square values 
in function fittings for yield ratios of 
cascades to pions as functions of multiplicity
for various $a_0$ parameters 
in Fig.~\ref{fig:a_0dependence}
and for ALICE data \cite{ALICE:2017jyt}
in the bottom line. 
}
\label{table:xipi-data}
\end{table}

\bibliography{inspire.bib}

\end{document}